\documentclass[prd,twocolumn,showpacs,showkeys,floatfix]{revtex4}
\usepackage{amsmath,amssymb}
\usepackage{latexsym}
\usepackage{bm}

\usepackage{hyperref}
\usepackage[ansinew]{inputenc}
\makeatletter
\def\eqnarray{\stepcounter{equation}\let\@currentlabel=\theequation
\global\@eqnswtrue
\global\@eqcnt\z@\tabskip\@centering\let\\=\@eqncr
$$\halign to \displaywidth\bgroup\@eqnsel\hskip\@centering
  $\displaystyle\tabskip\z@{##}$&\global\@eqcnt\@ne
  \hfil${\;##\;}$\hfil
  &\global\@eqcnt\tw@ $\displaystyle\tabskip\z@{##}$\hfil
   \tabskip\@centering&\llap{##}\tabskip\z@\cr}
\makeatother

\begin{document}
\title{Hyperbolically symmetric versions of Lemaitre--Tolman--Bondi spacetimes}
\author{L. Herrera}
\email{lherrera@usal.es}
\affiliation{Instituto Universitario de F\'isica
Fundamental y Matem\'aticas, Universidad de Salamanca, Salamanca 37007, Spain.}
\author{A. Di Prisco}
\email{alicia.diprisco@ciens.ucv.ve}
\affiliation{Escuela de F\'{\i}sica, Facultad de Ciencias,
Universidad Central de Venezuela, Caracas 1050, Venezuela.}
\author{J. Ospino}\affiliation{Departamento de Matem\'atica Aplicada and Instituto Universitario de F\'isica
Fundamental y Matem\'aticas, Universidad de Salamanca, Salamanca 37007, Spain}
\email{j.ospino@usal.es}
\begin{abstract}
We study fluid distributions endowed with hyperbolical symmetry, which share many common features with Lemaitre--Tolman--Bondi (LTB) solutions (e.g. they are geodesic, shearing, non--conformally flat and the energy density is inhomogeneous).  As such they may be considered as   hyperbolically symmetric versions of  LTB, with spherical symmetry replaced by hyperbolical symmetry. We start by considering pure dust models, and afterwards we extend our analysis  to dissipative models with anisotropic pressure. In the former case the complexity factor is necessarily non--vanishing, whereas in the latter cases models with vanishing complexity factor are found. The remarkable fact is that all solutions satisfying the vanishing complexity condition are necessarily non--dissipative and  satisfy the stiff equation of state.
\end{abstract}

\date{\today}
\maketitle
\section{INTRODUCTION}
In a recent paper  \cite{hd} we have presented a general study on the dynamics of hyperbolically symmetric fluids (DHSF).  Our main motivation behind such endeavor  (but not the only one) was to describe the dynamic regime preceding  the final equilibrium state of  static hyperbolically symmetric fluids described in \cite{st1}, and which in its turn could be used to model the source of the hyperbolically symmetric black hole described  in \cite{1h, 2h} where the region interior to the horizon is described by the line element
\begin{eqnarray}
ds^2&=&-\left(\frac{2M}{R}-1\right)dt^2+\frac{dR^2}{\left(\frac{2M}{R}-1\right)}+R^2d\Omega^2, \nonumber \\
d\Omega^2&=&d\theta^2+\sinh^2 \theta d\phi^2.
\label{w3}
\end{eqnarray}

The rationale behind  such a proposal  stems from    the well known fact that any transformation that maintains the static form of the
Schwarzschild metric (in the whole space--time) is unable to remove the coordinate singularity in the line element
\cite{rosen}. Or, in other words, the regular extension of the Schwarzschild metric to the whole space--time (including the region inner to the horizon) may be achieved but at the price to admit a non--static space--time inside the horizon \cite{Rin, Caroll}.

Since any dynamic regime should eventually end  in an equilibrium final state,
 it would be desirable  to have  a static solution over the whole space--time.

Thus the   model proposed in \cite{1h} describes the space time as
consisting of two four dimensional manifolds, one   described by  the usual  Schwarzschild metric on the exterior side of the horizon
and a second one  in the interior of it, described by (\ref{w3}).  

The metric (\ref{w3}) is a static solution admitting the four Killing vectors

\begin{equation}
\mathbf{\chi }_{(\mathbf{0})} = \partial _{\mathbf{t}},
\label{sest}
\end{equation}
and
\begin{eqnarray}
 {\bf \chi_{(2)}}=-\cos \phi
\partial_{\theta}+\coth\theta \sin\phi \partial_{\phi},\nonumber \\
{\bf \chi_{(1)}}=\partial_{\phi}, \quad {\bf \chi_{(3)}}=\sin \phi \partial_{\theta}+\coth\theta \cos\phi
\partial_{\phi}.
\label{shy}
\end{eqnarray}

The above Killing vectors (\ref{shy})  define the hyperbolic symmetry. Solutions to the Einstein equations  endowed with this type of symmetry  have been  the subject of research  in
different contexts (see \cite{Ha, ellis, 1n, Ga, Ri, mim, Ka, Ma, mimII} and references therein).

Besides the general properties  of DHSF analyzed in \cite{hd}, some exact solutions were found. In particular two non--dissipative solutions which could be regarded as the hyperbolically symmetric versions of the Friedman--Robertson--Walker space--time which  were analyzed in some detail. All solutions presented in \cite{hd} satisfy the condition of vanishing complexity factor, and evolve in the quasi--homologous regime. This last condition, in the non--dissipative case, implies that the fluid is shear--free, thereby excluding the possibility to obtain a hyperbolically symmetric version of the Lemaitre--Tolman--Bondi spacetimes (LTB).

Due to the huge relevance of LTB spacetimes, we shall devote this work to study in some detail its possible hyperbolically symmetric versions. For doing that we must abandon the condition of quasi--homologous evolution.

It is worth recalling  that LTB dust models \cite{1, 2, 3} are among the most appealing solutions to Einstein equations. They describe spherically symmetric distribution of   inhomogeneous non--dissipative dust (see \cite{4, 5} for a detailed description of these spacetimes). Although LTB space--times are usually associated with an inhomogeneous  dust source,  it is known  that  the most general source compatible with LTB space--times is an anisotropic fluid (\cite{4}, \cite{s1}).

LTB space--times have been invoked as cosmological models (see \cite{sn,  7, 6} and references therein), in the study of gravitational collapse,  when dealing with the problem of the cosmic censorship \cite{9, 10, 11, 12,  m1, m2}, and in quantum gravity \cite{13, 14}.

The apparent accelerated expansion of the universe, as inferred from  some observations of type Ia supernovae, has renewed the interest in LTB space--times.  Indeed, even though  there is  general  consensus to invoke dark energy as a source of anti-gravity for understanding
the cosmic acceleration, a growing number of researchers is now considering that  inhomogeneities may  account for the observed cosmic acceleration, without resorting to dark energy (see \cite{Coley1, Coley2,  15, 7'', celn, cel} and references therein).

In this work we shall present several models which could be considered as hyperbolically symmetric versions of LTB space--times. We shall consider both non--dissipative and dissipative models. The general approach used for reaching our goal was already outlined in \cite{hd}, however for sake of completeness we shall present the basic steps in the following sections. The last section is devoted to the discussion of the obtained results.

\section{FLUID DISTRIBUTION, KINEMATICAL VARIABLES AND BASIC EQUATIONS}
We consider  hyperbolically  symmetric distributions  of
geodesic fluid, which may be bounded (or not)  from outside by a surface $\Sigma^{e}$. As we already know  (see \cite{hd} for a detailed discussion on this point)  hyperbolically symmetric fluids cannot fill the central region and therefore such a region should be described either by an empty vacuole or by a fluid distribution not endowed with hyperbolical symmetry. Thus our   fluid is also bounded  from inside by a surface $\Sigma ^i$.

 Choosing comoving coordinates the general
 metric can be written as

\begin{equation}
ds^2=-dt^2+B^2dr^2+R^2(d\theta^2+\sinh^2\theta d\phi^2),
\label{1}
\end{equation}
where $B$ and $R$ are assumed
positive, and due to the symmetry defined by (\ref{shy}) are functions of $t$ and $r$. We number the coordinates $x^0=t$, $x^1=r$, $x^2=\theta$
and $x^3=\phi$.

The general energy momentum tensor $T_{\alpha\beta}$ of the fluid distribution
may be written as
\begin{eqnarray}
T_{\alpha\beta}&=&(\mu +
P_{\perp})V_{\alpha}V_{\beta}+P_{\perp}g_{\alpha\beta}+(P_r-P_{\perp})\chi_{
\alpha}\chi_{\beta}\nonumber \\&+&q_{\alpha}V_{\beta}+V_{\alpha}q_{\beta}
, \label{3}
\end{eqnarray}
where $\mu$ is the energy density, $P_r$ the radial pressure,
$P_{\perp}$ the tangential pressure and  $q^{\alpha}$ the heat flux, these physical variables, due to the symmetry defined by (\ref{shy}),  are  functions of $t$ and $r$.  Also, $V^{\alpha}$ and $\chi^{\alpha}$ denote four--velocity of the fluid and 
a unit four--vector along the radial direction respectively,  they
satisfy
\begin{eqnarray}
V^{\alpha}V_{\alpha}=-1, \;\; V^{\alpha}q_{\alpha}=0, \;\; \chi^{\alpha}\chi_{\alpha}=1,\;\;
\chi^{\alpha}V_{\alpha}=0.
\end{eqnarray}

Since we are considering comoving observers, we have
\begin{eqnarray}
V^{\alpha}=\delta_0^{\alpha}, \;\;
q^{\alpha}=qB^{-1}\delta^{\alpha}_1, \;\;
\chi^{\alpha}=B^{-1}\delta^{\alpha}_1.
\end{eqnarray}
\subsection{Einstein equations}
For (\ref{1}) and (\ref{3}), Einstein equations

\begin{equation}
G_{\alpha \beta} = 8 \pi T_{\alpha \beta}, \label{Eeq}
\end{equation}
read:
\begin{widetext}
\begin{eqnarray}
8\pi  \mu
=\left(2\frac{\dot{B}}{B}+\frac{\dot{R}}{R}\right)\frac{\dot{R}}{R}
-\left(\frac{1}{B}\right)^2\left[2\frac{R^{\prime\prime}}{R}+\left(\frac{R^{\prime}}{R}\right)^2
-2\frac{B^{\prime}}{B}\frac{R^{\prime}}{R}+\left(\frac{B}{R}\right)^2\right],
\label{12}
\end{eqnarray}
\end{widetext}
\begin{equation}
4\pi q=\frac{1}{B} \left(\frac{{\dot
R}^{\prime}}{R} -\frac{\dot B}{B}\frac{R^{\prime}}{R}\right),
\label{13}
\end{equation}
\begin{eqnarray}
8\pi P_r
=-\left[2\frac{\ddot{R}}{R}+\left(\frac{\dot{R}}{R}\right)^2\right]
+\left(\frac{R^{\prime}}{BR}\right)^2+\left(\frac{1}{R}\right)^2,
\label{14}
\end{eqnarray}
\begin{equation}
8\pi P_\bot
=-\left(\frac{\ddot{B}}{B}+\frac{\ddot{R}}{R}
+\frac{\dot{B}}{B}\frac{\dot{R}}{R}\right)
+\left(\frac{1}{B}\right)^2\left(
\frac{R^{\prime\prime}}{R}
-\frac{B^{\prime}}{B}\frac{R^{\prime}}{R}\right),\label{15}
\end{equation}
where dots and primes denote derivatives with respect to $t$ and $r$ respectively. It is worth stressing the difference between these equations and the corresponding to the spherically symmetric  LTB case.

\subsection{Kinematical variables and the mass function}

The expansion $\Theta$ is given by

\begin{equation}
\Theta={V^{\alpha}}_{;\alpha}=\left(\frac{\dot{B}}{B}+2\frac{\dot{R}}{R}\right),\label{th}
\end{equation}
and for the shear we have (remember that the four--acceleration and the vorticity vanish)

\begin{equation}
\sigma_{\alpha\beta}=V_{(\alpha
;\beta)}-\frac{1}{3}\Theta h_{\alpha\beta}, \label{4a}
\end{equation}
where $h_{\alpha \beta} = g_{\alpha \beta} + V_\alpha V_\beta $.

The non--vanishing components
of  (\ref{4a}) are
\begin{equation}
\sigma_{11}=\frac{2}{3}B^2\sigma, \;\;
\sigma_{22}=\frac{\sigma_{33}}{\sinh^2\theta}=-\frac{1}{3}R^2\sigma,
 \label{5a}
\end{equation}
with
\begin{equation}
\sigma^{\alpha\beta}\sigma_{\alpha\beta}=\frac{2}{3}\sigma^2,
\label{5b}
\end{equation}
being
\begin{equation}
\sigma=\left(\frac{\dot{B}}{B}-\frac{\dot{R}}{R}\right).\label{5b1}
\end{equation}
$\sigma_{\alpha\beta}$ may be  also  written as
\begin{equation}
\sigma_{\alpha \beta}= \sigma \left(\chi_\alpha \chi_\beta -
\frac{1}{3} h_{\alpha \beta}\right). \label{sh}
\end{equation}

Next, the mass function $m(t,r)$ introduced by Misner and Sharp \cite{Misner} (see also \cite{Cahill})
is given by

\begin{equation}
m=-\frac{R}{2}R^{3}_{232}=\frac{R}{2}\left[-\dot R^2+\left(\frac{R^{\prime}}{B}\right)^2+1 \right],
\label{18}
\end{equation}
where the components $R^{3}_{232}$ of the Riemann tensor is  calculated with (\ref{1}).

Defining as usual  the ``areal''  velocity $U$ of the 
fluid as the variation of $R$ with respect to proper time, i.e.\
\begin{equation}
U=\dot R, \label{19}
\end{equation}
then since $U<1$ it follows at once from (\ref{18}) that $m$ is a positive defined quantity.

With the above  we can express (\ref{18}) as
\begin{equation}
E\equiv \frac{R^\prime}{B}=\left (U^2+\frac{2 m}{R}-1\right)^{1/2}.\label{21a}
\end{equation}

From
(\ref{18}) and field equations we obtain
\begin{eqnarray}
\dot m=4\pi R^2\left(P_r U+q E\right),
\label{22}
\end{eqnarray}
and
\begin{eqnarray}
m^{\prime}=-4\pi R^{\prime}R^2 \left(\mu +q \frac{U}{E}\right).
\label{27}
\end{eqnarray}
The integration of  (\ref{27})  produces
\begin{equation}
m=-\int^{r}_{0}4\pi R^2 \left(\mu+q \frac{U}{E}\right)
R^{\prime}dr,\label{27int}
\end{equation}
whose partial integration yields
\begin{equation}
\frac{3m}{R^3} = -4\pi \mu +\frac{4\pi}{R^3} \int^r_0{R^3 \left(\mu^{\prime}-3 q\frac{UB}{R}\right)dr}.
\label{3m/R3}
\end{equation}
Then,  it follows from (\ref{27int}) that $\mu$ is necessarily negative, if we assume   the condition $R^\prime>0$ to avoid shell crossing,  and remind that $m>0$.

 Furthermore, it follows from (\ref{27int}) that whenever the energy density  is regular, then $m\sim r^3$ as $r$ tends to zero. However, in this same limit $U\sim 0$, and $R\sim r$ implying because of (\ref{21a}) that the central region  cannot be filled with our fluid distribution. Among the many  possible scenarios we shall assume here that the center is surrounded by a vacuum cavity. However,  this is just one of the possible choices, which even if having implications on  specific models, does not affect  the general properties of the  fluids endowed with hyperbolical symmetry.
\subsection{The exterior spacetime and junction conditions}
In the case of bounded configurations, we assume that outside $\Sigma^{e}$, we have the hyperbolic symmetric version of the Vaidya spacetime,
described by:
\begin{equation}
ds^2=-\left[\frac{2M(v)}{\rho}-1\right]dv^2-2d\rho dv+\rho^2(d\theta^2
+\sinh^2\theta
d\phi^2) \label{1int},
\end{equation}
where $M(v)$ denotes the total mass and $v$ is the retarded~time.

Now, from~the continuity of the first differential form, it follows (see~\cite{chan1} for details),
\begin{equation}
 dt\stackrel{\Sigma^{e}}{=}dv \left(\frac{2M(v)}{\rho}-1\right), \label{junction1f}
\end{equation}
\begin{equation}
R\stackrel{\Sigma^{e}}{=}\rho(v), \label{junction1f2}
\end{equation}
and:
\begin{equation}
\left(\frac{dv}{dt}\right)^{-2}\stackrel{\Sigma^{e}}{=}\left(\frac{2M(v)}{\rho}-1+2\frac{d\rho}{dv}\right),\label{junction1f3}
\end{equation}

whereas the continuity of the second differential form produces:
\begin{equation}
m(t,r)\stackrel{\Sigma^{e}}{=}M(v), \label{junction1}
\end{equation}
and:
\begin{equation}
P_r\stackrel{\Sigma^{e}}{=}q, \label{junction1}
\end{equation}
where $\stackrel{\Sigma^{e}}{=}$ means that both sides of the equation are evaluated on $\Sigma^e$.

The corresponding junction conditions on $\Sigma^i$ are:
\begin{equation}
m(t,r)\stackrel{\Sigma^{i}}{=}0, \label{junction1}
\end{equation}
and:
\begin{equation}
P_r\stackrel{\Sigma^{i}}{=}0. \label{junction1}
\end{equation}

When either of the above conditions cannot be satisfied, we have to admit the presence of thin~shells.

\subsection{ Weyl tensor}

The Weyl tensor is defined through the  Riemann tensor
$R^{\rho}_{\alpha \beta \mu}$, the  Ricci tensor
$R_{\alpha\beta}$ and the curvature scalar $\cal R$, as:
$$
C^{\rho}_{\alpha \beta \mu}=R^\rho_{\alpha \beta \mu}-\frac{1}{2}
R^\rho_{\beta}g_{\alpha \mu}+\frac{1}{2}R_{\alpha \beta}\delta
^\rho_{\mu}-\frac{1}{2}R_{\alpha \mu}\delta^\rho_\beta$$
\begin{equation}
+\frac{1}{2}R^\rho_\mu g_{\alpha \beta}+\frac{1}{6}{\cal
R}(\delta^\rho_\beta g_{\alpha \mu}-g_{\alpha
\beta}\delta^\rho_\mu). \label{34}
\end{equation}

In our case the magnetic part of the Weyl tensor vanishes, whereas its electric  part, defined by
\begin{equation}
E_{\alpha \beta} = C_{\alpha \mu \beta \nu} V^\mu V^\nu,
\label{elec}
\end{equation}
has  the following non--vanishing components
\begin{eqnarray}
E_{11}&=&\frac{2}{3}B^2 {\cal E},\nonumber \\
E_{22}&=&-\frac{1}{3} R^2 {\cal E}, \nonumber \\
E_{33}&=& E_{22} \sinh^2{\theta}, \label{ecomp}
\end{eqnarray}
where
\begin{eqnarray}
&&{\cal E}= \frac{1}{2}\left[\frac{\ddot R}{R} - \frac{\ddot B}{B} - \left(\frac{\dot R}{R} - \frac{\dot B}{B}\right) \frac{\dot R}{R}\right]\nonumber \\
&+& \frac{1}{2 B^2} \left[ -
\frac{R^{\prime\prime}}{R} + \left(\frac{B^{\prime}}{B} +
\frac{R^{\prime}}{R}\right) \frac{R^{\prime}}{R}\right]
+\frac{1}{2 R^2}. \label{E}
\end{eqnarray}

Observe that we may also write $E_{\alpha\beta}$ as:
\begin{equation}
E_{\alpha \beta}={\cal E} (\chi_\alpha
\chi_\beta-\frac{1}{3}h_{\alpha \beta}). \label{52}
\end{equation}

Finally, using (\ref{12}), (\ref{14}), (\ref{15}) with (\ref{18}) and (\ref{E}) we obtain
\begin{equation}
\frac{3m}{R^3}=-4\pi \mu+4\pi(P_r-P_\bot) +\cal{E}.
\label{mE}
\end{equation}

\section{Structure Scalars and Complexity Factor}
Some of the models exhibited below are obtained from the conditions imposed on a scalar function that appears in a natural way in the orthogonal splitting of the Riemann tensor (see~\cite{sp} for details) and~that is identified as the complexity~factor. 

Thus, let us introduce the tensor $Y_{\alpha \beta}$ (which is an element of that splitting~\cite{16b, 17b, 18b, parrado}), defined
by:
\begin{equation}
Y_{\alpha \beta}=R_{\alpha \gamma \beta \delta}V^\gamma V^\delta.
\label{electric}
\end{equation}

Tensor $Y_{\alpha \beta}$ may be expressed in terms of two scalar functions $Y_T, Y_{TF}$ (structure scalars) as:
\begin{eqnarray}
Y_{\alpha\beta}=\frac{1}{3}Y_T h_{\alpha
\beta}+Y_{TF}(\chi_{\alpha} \chi_{\beta}-\frac{1}{3}h_{\alpha
\beta}),\label{electric'}
\end{eqnarray}
where:
\begin{eqnarray}
Y_T=4\pi(\mu+3 P_r-2\Pi) , \qquad
Y_{TF}={\cal E}-4\pi \Pi ,\label{EY}
\end{eqnarray}
with $\Pi=P_r-P_\bot$.

Combining (\ref{mE}) with (\ref{3m/R3}) and (\ref{EY}) produces:
\begin{equation}\label{YTF}
 Y_{TF}=-8\pi \Pi+\frac{4\pi}{R^3} \int^r_0{R^3\left(\mu^\prime -3 q \frac{UB}{R}\right) dr}.
\end{equation}

The complexity factor is a scalar function intended to measure the complexity of a given fluid distribution (see~\cite{c1, c2, epjc} for details). For~static hyperbolic symmetric fluids (as well as for spherically symmetric ones), the complexity factor is identified with the scalar function $Y_{TF}$ defined, in~the dynamic case, by~Equations~(\ref{EY}) and (\ref{YTF}) (see~\cite{hd}).
The main reason behind such a proposal resides, on~the one hand, in~the basic assumption that one of the less complex systems corresponds to a homogeneous (in the energy density) fluid distribution with isotropic pressure. Thus, any variable measuring complexity should vanish for this specific case. On~the other hand, the~scalar function $Y_{TF}$ contains contributions from the energy density inhomogeneity and the local pressure anisotropy, combined in a very specific way, which (in the static case) vanishes for the homogeneous and locally isotropic fluid distribution. Furthermore, this scalar measures the departure of the value of the Tolman mass for the homogeneous and isotropic fluid, produced by the energy density inhomogeneity and the pressure~anisotropy.

It is worth mentioning that the complexity factor so defined not only vanishes for the simple configuration mentioned above, but also may vanish when the terms appearing in its definition cancel each other. Thus, vanishing complexity may correspond to very different~systems.

In the time-dependent case, we face two different problems: on the one hand, we have to generalize the concept of the complexity of the structure of the fluid distribution to time-dependent dissipative fluids, and~on the other hand, we also have to evaluate the complexity of the mode of evolution. Following the strategy outlined in~\cite{c2}, the~complexity factor for the dissipative case of the fluid distribution is assumed to be the function $Y_{TF}$, as~in the static case, which now includes the dissipative variables. With~respect to the complexity of the mode of evolution, let us recall that in the past, the homologous and quasi-homologous conditions have been used to characterize the simplest mode of evolution. However, we know that in the nondissipative case, the homologous and the quasi-homologous conditions imply the vanishing of the shear (see Equation~(59) in~\cite{hd}), and~therefore, we shall not adopt such restrictions~here.

\section{Hyperbolically symmetric Lemaitre--Tolman--Bondi metric: The non--dissipative dust case}
We start our search of hyperbolically symmetric  versions of LTB (HSLTB)  by considering the simplest case, i.e. we  assume  non-- dissipative geodesic dust.  Extensions  to the dissipative, anisotropic   case  shall be discussed in the next section, along the lines developed in \cite{LTBd}.

Under the above mentioned conditions, 
we find after integration of  (\ref{13})

\begin{equation}
B(t,r)=\frac{R^\prime}{\left[k(r)-1\right]^{1/2}},\label{BTBh}
\end{equation}
\noindent where $k$ is an arbitrary function of $r$.

 Then from (\ref{18}) and  (\ref{BTBh}) it follows
\begin{equation}\dot R^2=-\frac{2m}{R}+k(r).
\label{intltb1h}
\end{equation}

Equation  (\ref{intltb1h}) implies  $k(r)>\frac{2m}{R}$. Thus unlike the spherically symmetric LTB space--time we now have only one case $k(r)>0$.

The solution to (\ref{intltb1h}) may be written as:

\begin{equation}
 R=\frac{m}{k}(\cosh \eta+1), \qquad  \frac{m}{k^{3/2}}(\sinh \eta+\eta)=t-t_{0}(r),
\label{int5h}
\end{equation}
where $t_{0}(r)$ is an integration function of $r$.

Thus, for the line element we have
\begin{equation}
ds^2=-dt^2+\frac{(R^{\prime})^2}{k(r)-1}dr^2+R^2(d\theta^2+\sinh^2\theta
d\phi^2).\label{mTB}
\end{equation}

In order to prescribe an explicit model we have to provide the three functions $k(r)$, $m(r)$ and $t_{0}(r)$. However, since (\ref{mTB}) is invariant under  transformations of the form $r=r(\tilde r)$, we only need two functions of $r$.

Assuming  $m_0=\frac{m}{k}=constant$, and $t_{0}(r)=constant$ the expressions for $\Theta$ and  $\sigma$ read
\begin{eqnarray}
  \Theta &=& \frac{\sqrt{k}}{m_0}\left ( \frac{\sinh \eta}{\sinh \eta +\eta}+\frac{\cosh \eta+2\sinh\eta+1}{(\cosh \eta+1)^2} \right ), \\
  \sigma &=& \frac{\sqrt{k}}{m_0}\left ( \frac{\sinh \eta}{\sinh \eta +\eta}+\frac{\cosh \eta-\sinh\eta+1}{(\cosh \eta+1)^2} \right ),
\end{eqnarray}
from where it is clear that the expansion is always positive.

Since, as we have already mentioned, our fluid distribution cannot reach the central region, then  we do not need to consider any regularity conditions there.

The only non--trivial conservation law in this case reads
\begin{equation}
\dot \mu+\mu \Theta=0,
\label{cl1}
\end{equation}
or

\begin{eqnarray}
\dot \mu+\mu\left(\frac{\dot B}{B}+2\frac{\dot R}{R}\right)=0,\label{an'}
\end{eqnarray}
producing
\begin{equation}
 \mu=\frac{h(r)}{BR^2},\label{nltb}
\end{equation}
or, using (\ref{BTBh})
\begin{equation}
 \mu=\frac{3 h(r)\left[k(r)-1\right]^{1/2}}{(R^{3})^\prime},\label{n1ltb}
\end{equation}
where $h(r)$ is a function of integration, which due to the fact that the energy density is negative, must be necessarily negative.

Scalar
 $Y_{TF}$  for (\ref{mTB}) reads 
\begin{equation}
Y_{TF}=\frac{\ddot R}{R}-\frac{\ddot R^\prime}{R^\prime}
.\label{EYTB}
\end{equation}

As it is evident from  (\ref{YTF}), since we are considering non--dissipative inhomogeneous dust, the complexity factor $Y_{TF}$ cannot vanish. However, this situation may change in the dissipative, anisotropic  case as we shall see in the models exhibited below.  On the other hand  since in the non--dissipative case the quasi--homologous condition implies the vanishing of the shear (see Eq.(59) in \cite{hd}), we have to abandon such a restriction for our models.

\section{Dissipative case}
We shall now consider the possibility that the system radiates, and the pressure is non--vanishing and may be anisotropic. For doing so, following the approach  presented in \cite{LTBd},  let us assume

\begin{equation}
B(t,r)=\frac{R^\prime}{\sqrt{K(t,r)-1}},\label{ecB}
\end{equation}
then integrating (\ref{13}) we find
\begin{equation}\label{K1}
  K(t,r)-1=\left [ \int 4\pi {q}Rdt  +C(r) \right ]^2,
\end{equation}
since in the non-dissipative case (\ref{ecB}) becomes (\ref{BTBh}), then $C(r)=\sqrt{k(r)-1})$.

\noindent Thus the line element reads

\begin{equation}
  ds^2=-dt^2+\frac{(R^\prime)^2 dr^2}{\left [ \int 4\pi {q}Rdt  +C(r) \right ]^2}+R^2(d\theta^2 +\sinh ^2\theta d\phi^2).
\end{equation}

Since we are considering  dissipative systems we shall need a transport equation. For simplicity we shall adopt here the transport equation ensuing from the so called ``truncated''  theory  \cite{19n}, it reads 
\begin{equation}
\tau
h^{\alpha\beta}V^{\gamma}q_{\beta;\gamma}+q^{\alpha}=-\kappa h^{\alpha\beta}
T_{,\beta} \label{V1},
\end{equation}
whose    only non--vanishing independent component becomes
\begin{equation}
\tau \dot q+q=-\frac{\kappa}{B}T^{\prime}, \label{V2}
\end{equation}
where $\kappa$ and $\tau$ denotes the thermal conductivity and the relaxation time respectively.

In order to obtain specific models we shall need to impose additional conditions. A first family of models will be obtained from conditions on the complexity factor, while a second family will be obtained by a specific restriction on the function $B$, particularly suitable for describing situations  where a cavity surrounding the central region appears.

\subsection{Models obtained upon conditions on the complexity factor}

In our case the complexity factor $Y_{TF}$ may be written as 
\begin{eqnarray}
  Y_{TF} &=& \frac{\ddot{R}}{R}-\frac{\ddot{R}^\prime}{R^\prime} +\frac{\ddot{K}}{2(K-1)}\nonumber \\
  &+&\frac{\dot{K}}{K-1}\left (\frac{\dot{R}^\prime}{R^\prime}-\frac{3}{4}\frac{\dot K}{K-1} \right).
\end{eqnarray}
\noindent In order to obtain models we shall first assume that the above structure scalar has the same form as in the non-dissipative case, implying
\begin{equation}
\frac{\ddot{K}}{2(K-1)}
  +\frac{\dot{K}}{K-1}\left (\frac{\dot{R}^\prime}{R^\prime}-\frac{3}{4}\frac{\dot K}{K-1} \right)=0.\label{ecK}
\end{equation}

\noindent The integration of  (\ref{ecK}) produces
\begin{equation}
\frac{R^\prime \sqrt{\dot{K}}}{(K-1)^{\frac{3}{4}}}=C_1(r)\label{pik},
\end{equation}
where $C_1$ is an integration function.
\noindent Integrating the above equation we obtain

\begin{equation}
K-1=\frac{4}{[-C_1^2\int \frac{dt}{(R^\prime) ^2}+C_2(r)]^2}\label{ecK2},
\end{equation}
where $C_2$ is another integration function.

\noindent Combining  (\ref{ecK2}) and  (\ref{K1}) it follows that
\begin{equation}\label{ecK3}
  \int 4\pi {q}Rdt  +C(r)  =\frac{2}{-C_1^2\int \frac{dt}{(R^\prime) ^2}+C_2(r)},
\end{equation}

\noindent or

\begin{equation}
C_2(r)=\frac{2}{\int 4\pi  qR dt+C(r)}+C_1^2\int \frac{dt}{(R^\prime)^2}.
\end{equation}
\noindent Using (\ref{pik}) it follows that  $C_2(r)=0$, thus
\begin{equation}
\frac{2}{\int 4\pi  qR dt+C(r)}+C_1^2\int \frac{dt}{(R^\prime)^2}=0,
\end{equation}

\noindent implying that  (\ref{ecK3}) may be written as
\begin{equation}\label{ecK4}
  2\pi {q}=\frac{1}{R\left ( R^\prime C_1(r) \int \frac{dt}{(R^\prime)^2}\right)^2}.
\end{equation}

Let us first try to obtain models of dissipative dust  satisfying (\ref{ecK2}).

\noindent Then using  (\ref{ecB}) and  (\ref{ecK}) we obtain

\begin{equation}
\frac{\dot{B}}{B}=\frac{\dot{R}^\prime}{R^\prime}-\frac{\dot{K}}{2(K-1)},\qquad \frac{\ddot{B}}{B}=\frac{\ddot{R}^\prime}{R^\prime}.\label{Bder}
\end{equation}

\noindent Feeding back  (\ref{Bder}) in (\ref{14})-(\ref{15}) produces
\begin{eqnarray}
  K &=&2\ddot{R}R+\dot{R}^2 \label{K11},\\
  K^\prime &=& 2R^\prime R\left [\frac{\ddot{R}^\prime}{R^\prime}+\frac{\ddot{R}}{R}+\frac{\dot{R}}{R}
  \left (\frac{\dot{R}^\prime}{R^\prime}-\frac{C_1^2\sqrt{K-1}}{2(R^\prime)^2}\right )\right]\label{K22}.
\end{eqnarray}

\noindent Next, taking the $r$ derivative  of (\ref{K11}) and replacing it in (\ref{K22}) we find

\begin{equation}\label{K1122}
 \frac{\dot{R}}{R} \frac{C_1^2\sqrt{K-1}}{2(R^\prime)^2}=0,
\end{equation}

\noindent from which it follows at once that there are no radiating dust solutions in this case. Therefore in the following subsection we shall relax the dust condition, and we shall consider models of radiating anisotropic fluids.

\subsubsection{ $P_\bot=0$,  $P_r\neq 0$ }

Let us consider models with vanishing  tangential pressure. 

Replacing  (\ref{ecB}) in (\ref{15}) the following expression for $P_\bot$ is found
\begin{equation}\label{Pbot0}
  8\pi P_\bot=-\left [\frac{\ddot{B}}{B} +\frac{\ddot{R}}{R}+\frac{\dot{R}}{R}\left(\frac{\dot{R}^\prime}{R^\prime}-\frac{\dot{K}}{2(K-1)}\right)\right]+\frac{K^\prime}{2RR^\prime}.
\end{equation}

In order to obtain a model, let us choose
\begin{equation}\label{Ele}
 \frac{\dot{R}}{R} \frac{\dot{R}^\prime}{R^\prime}=\frac{K^\prime}{2RR^\prime},
\end{equation}
implying
\begin{equation}
 K-1=\dot{R}^2. \label{ahs}
\end{equation}

 Then, from the condition $P_\bot=0$, we obtain from (\ref{Pbot0})

\begin{equation}\label{BPbot}
  \ddot{B}=0,\quad \Rightarrow B=b_1(r) t+b_2(r),
\end{equation}
  where $b_1$ and $b_2$ are two arbitrary functions.

Using  (\ref{ecB}) we find for  $R$
\begin{equation}\label{RPbot}
  R^\prime-B\dot{R}=0, \quad \Rightarrow\quad R=\Phi\left[a_1(r) t+a_2(r)\right],
\end{equation}
where $\Phi$ is an arbitary function of its argument
\noindent and
\begin{equation}
a_1(r)=e^{\int b_1(r)dr},\qquad a_2(r)=\int b_2(r)e^{\int b_1(r)dr}dr.\label{a1a2}
\end{equation}

Then the physical variables read
\begin{eqnarray}
  8\pi \mu &=&- \frac{1}{R^2}-\frac{\dot{R}}{R}\frac{\dot{K}}{(K-1)} ,\\
  4\pi q &=& \frac{\dot{K}}{2R\sqrt{K-1}}, \\
  8\pi P_r &=& -\frac{\dot{R}}{R}\frac{\dot{K}}{K-1}+\frac{1}{R^2}.
\end{eqnarray}

For specifying further the model let us choose  $b_1(r)$ and $b_2(r)$ as

\begin{equation}\label{bb1}
  b_1(r)=\frac{\beta_1}{r+\beta_2},\qquad b_2(r)=(r+\beta_2)^\alpha,
\end{equation}
implying
\begin{equation}\label{aa2}
  a_1(r)=(r+\beta_2)^{\beta_1},\qquad a_2(r)=\frac{(r+\beta_2)^{\alpha+\beta_1+1}}{\alpha+\beta_1+1},
\end{equation}
and
\begin{eqnarray}
  B &=& \frac{\beta_1 t}{r+\beta_2}+(r+\beta_2)^{\alpha},\\
  R &=& (a_1t+a_2)^n,
\end{eqnarray}
where $\beta_1$, $\beta_2$, $\alpha$ and $n$ are arbitrary constants.

\noindent Thus, the physical and kinematical variables for this model read
\begin{eqnarray}
  8\pi \mu &=&-\frac{1}{(a_1t+a_2)^{2n}} -\frac{2(n-1)na_1^2}{(a_1t+a_2)^2},\label{d2}\\
  4\pi q &=& \frac{(n-1)na_1^2}{(a_1t+a_2)^2},\label{qn1} \\
  8\pi P_r &=&\frac{1}{(a_1t+a_2)^{2n}} -\frac{2(n-1)na_1^2}{(a_1t+a_2)^2},\label{p2}\\
m&=&\frac{(a_1t+a_2)^n}{2},
\end{eqnarray}

\begin{eqnarray}
  \Theta &=& \frac{b_1}{b_1t+b_2}+\frac{2n a_1}{a_1t+a_2} ,\\
 \sigma&=& \frac{b_1}{b_1t+b_2}-\frac{n a_1}{a_1t+a_2}.
\end{eqnarray}

A simple calculation of  the complexity factor ($Y_{TF}$) for this model produces
\begin{equation}
Y_{TF}=\frac{(n-1)na_1^2}{(a_1t+a_2)^2}.
\label{cf1}
\end{equation}
It is worth noticing that it has exactly the same expression as $q$  as given by (\ref{qn1}). Therefore any solution of this family satisfying the vanishing complexity factor is necessarily non--dissipative.
On the other hand  $Y_{TF}$  is zero  if $n=0$ and/or $a_1=0$ and/or $n=1$. The first two conditions are ruled out at once from (\ref{aa2}) and (\ref{d2}). Thus the solution of this family with vanishing complexity factor is characterized by $n=1$, which using (\ref{d2}) and (\ref{p2}) produces
\begin{equation}
P_r=-\mu.
\label{st}
\end{equation}
The above is the stiff equation of state originally  considered by Zeldovich (see \cite{c1}).

\subsection{Models with $B=1$}.

We shall next assume $B=1$ in order to obtain some analytical models. As discussed in \cite{16n}, such a condition is particularly suitable for describing fluid distributions whose center is surrounded by an empty cavity, a scenario we expect for the kind of fluid distributions we are dealing with in this work.

The corresponding Einstein equations may be written as
\begin{eqnarray}\label{EEB1}
 8\pi  \mu &=& -\frac{1}{R^2}-\frac{2 R^{\prime\prime}}{R}-\left(\frac{R^\prime}{R}\right)^2+\frac{\dot{R}^2}{R^2}, \\
  4\pi q &=& \frac{\dot{R}^\prime}{R}, \label{101}\\
  8\pi  P_r &=& \frac{1}{R^2} +\left(\frac{R^\prime}{R}\right)^2-\left [\left(\frac{\dot R}{R}\right)^2+\frac{2\ddot{R}}{R}\right],\label{102}\\
  8\pi P_\bot &=&  \frac{R^{\prime\prime}}{R}-\frac{\ddot{R}}{R}.\label{103}
\end{eqnarray}
\subsubsection{Non dissipative case}
Let us first consider  the non--dissipative case ($q=0$). In this case it follows at once from (\ref{101})  that $R$ is a separable function, i.e. it takes the form

\begin{equation}
R=R_1(t)+R_2(r),
\label{se1}
\end{equation}
where $R_1$ and $R_2$ are arbitrary functions of their  arguments.

Using (\ref{se1}) in (\ref{BTBh}) it follows at once that
\begin{equation}
R_2^\prime=\sqrt{k(r) -1}.
\label{se2}
\end{equation}

In order to exhibit an exact solution let us further assume $P_\bot=0$. Using this condition in (\ref{103}) produces
\begin{equation}
R_1(t)=a t^2+b_1t +c_1,\qquad R_2(r)=a r^2+b_2 r+c_2,
\label{se3}
\end{equation}
where $a, b_1, c_1, b_2, c_2$ are arbitrary constants.

The physical and kinematical variables for this model are
\begin{equation}
8\pi \mu=\frac{1}{\alpha^2}\left(-1-4a \alpha-\beta^2+\gamma^2\right),
\label{se4}
\end{equation}

\begin{equation}
8\pi P_r=\frac{1}{\alpha^2}\left(1-4a \alpha+\beta^2-\gamma^2\right),
\label{se5}
\end{equation}

\begin{equation}
P_r+\mu=-\frac{a}{\pi \alpha},
\label{semp}
\end{equation}

\begin{equation}
\Theta=\frac{2 \gamma}{\alpha}\label{se6},
\end{equation}

\begin{equation}
\sigma=-\frac{\gamma}{\alpha}\label{se7},
\end{equation}

\begin{equation}
m=\frac{ \alpha}{2}\left(\beta^2-\gamma^2+1\right),\label{se8}
\end{equation}
where
\begin{eqnarray}
\alpha&\equiv& a(t^2+r^2)+b_1t+b_2r+c_1+c_2;\qquad \beta\equiv 2ar+b_2, \nonumber \\ &&\gamma\equiv 2at+b_1.
\label{se9}
\end{eqnarray}

For this model the expression for $Y_{TF}$ reads
\begin{equation}
Y_{TF}=\frac{2a}{\alpha}.
\label{cf3}
\end{equation}

Therefore the vanishing complexity factor implies $a=0$, producing because  of (\ref{se4})  and (\ref{se5})
\begin{equation}
P_r=-\mu.
\label{st3}
\end{equation}
Thus the solution of this family with the vanishing complexity factor  condition is also characterized by the stiff equation of state.
\subsubsection{Dissipative case}
Let us now consider the dissipative case ($q\neq0$). If we impose the condition $P_\bot=0$, then we get the equation $\ddot R=R^{\prime \prime}$, whose general solution is of the form
\begin{equation}
R(t,r)=c_1 \Psi(t+r) +c_2\Phi(t-r),
\label{se10}
\end{equation}
where $c_1, c_2$ are arbitrary constants, and $\Psi, \Phi$ arbitrary functions of their arguments.

As an example let us choose
\begin{equation}
R(t,r)=c \sin a(t-r),
\label{se11}
\end{equation}
where $a, c$ are arbitrary constants. Then,  for the kinematical and physical variables we obtain
\begin{equation}
8\pi \mu=2 a^2-\frac{1}{c^2\sin^2[a(t-r)]},
\label{se12}
\end{equation}
\begin{equation}
4\pi q=a^2,
\label{se13}
\end{equation}
\begin{equation}
8\pi P_r=2 a^2+\frac{1}{c^2\sin^2[a(t-r)]},
\label{se14}
\end{equation}
\begin{equation}
\Theta=2 a \cot[a(t-r)],
\label{se15}
\end{equation}
\begin{equation}
\sigma=-a \cot[a(t-r)],
\label{se16}
\end{equation}

\begin{equation}
m=\frac{c \sin [a(t-r)]}{2},
\label{se17}
\end{equation}

\noindent For this case the temperature  $T(t,r)$, calculated from  (\ref{V2}), reads

\begin{equation}\label{T1}
  T(t,r)=-\frac{a^2 r}{4\pi \kappa}+T_0(t),
\end{equation}

whereas the expression for  $Y_{TF}$ is
\begin{equation}
Y_{TF}=-a^2,
\label{cf4}
\end{equation}
implying because of (\ref{se11}) that no solution of this family has a vanishing complexity factor.

Finally, as an alternative model we may assume
\begin{equation}
R=a(t-r)^n,
\label{se18}
\end{equation}
where $a, n$ are constants.

The ensuing physical and kinematical variables are in this case:
\begin{equation}
8\pi \mu=-\frac{1}{a^2\left(t-r\right)^{2n}}-\frac{2n\left(n-1\right)}{\left(t-r\right)^2}
\label{se19}
\end{equation}
\begin{equation}
4\pi q=-\frac{n\left(n-1\right)}{\left(t-r\right)^2},
\label{se20}
\end{equation}
\begin{equation}
8\pi P_r=\frac{1}{a^2\left(t-r\right)^{2n}}-\frac{2n\left(n-1\right)}{\left(t-r\right)^2},
\label{se21}
\end{equation}
\begin{equation}
\Theta=\frac{2 n}{t-r},
\label{se22}
\end{equation}
\begin{equation}
\sigma=-\frac{ n}{t-r},
\label{se23}
\end{equation}

\begin{equation}
m=\frac{a\left(t-r\right)^n}2.
\label{se24}
\end{equation}

\noindent Using (\ref{V2}), the expression for  the temperature becomes

\begin{equation}\label{T2}
  T(t,r)=\frac{n(n-1)}{4\pi \kappa (t-r)}-\frac{n(n-1)\tau}{4\pi \kappa (t-r)^2}+T_0(t).
\end{equation}

The complexity factor for this family of solutions reads
\begin{equation}
Y_{TF}=\frac{n(n-1)}{(t-r)^2}.
\label{cf5}
\end{equation}

The above scalar may vanish only if $n=0$ and/or $n=1$. The first possibility has to be ruled out from a simple inspection on (\ref{se19}), and therefore the vanishing complexity factor  conditions requires $n=1$, implying because of (\ref{se20}) that the fluid is non--dissipative, and because of (\ref{se19}) and (\ref{se21}) that the fluid satisfies the stiff equation of state $P_r=-\mu$.

\section{CONCLUSIONS}

We have investigated in some detail all possible solutions of fluids endowed with the hyperbolical symmetry (\ref{shy}), characterized by non--vanishing shear, inhomogeneous energy--density and vanishing four--acceleration (geodesics). So defined, these solutions are entitled to be considered as  hyperbolically symmetric versions of LTB space--times.

The first class of solutions corresponds to non--dissipative dust configurations. Comparing with the spherically symmetric case we observe that only one family of solutions   ($k(r)>0$) exists, instead of the three  families existing in this latter case ($k(r)\lesseqqgtr 0$). 

These solutions cannot satisfy the vanishing complexity factor, neither can they evolve in the quasi--homologous regime. On the other hand the scalar expansion is positive as expected for pure dust submitted to a repulsive gravity.

Next we have analyzed the case of dissipative anisotropic fluids. For doing this we generalized the expression (\ref{BTBh})  by assuming (\ref{ecB}). Different specific models were found from two different conditions. One class of solutions was obtained from a condition imposed on the complexity factor (\ref{ecK}). It was shown that in this case   the pressure must be anisotropic. A solution of this type was found assuming further that $P_\bot=0$. The subclass of this solution satisfying the vanishing complexity factor is necessarily non--dissipative, and satisfies the stiff equation of state $P_r=-\mu$.

The other class of solutions was found under the condition $B=1$. For the non--dissipative case a family of solutions was found under the additional condition $P_\bot=0$. In this case too, the vanishing complexity factor condition implies the stiff equation of state $P_r=-\mu$. In the dissipative case two families of solutions were found from different assumptions on the specific form of $R$. Thus, assuming (\ref{se11}) we found a solution  never satisfying the vanishing complexity factor condition, whereas assuming (\ref{se18}) such a condition can be satisfied implying that the fluid is non--dissipative and satisfies the  stiff equation of state $P_r=-\mu$. It is worth noticing that the temperature for the first of the above solutions (\ref{T1}), does not contain terms depending on the relaxation time. In other words the model behaves as if  the fluid is always in the thermal stationary state, a result that becomes intelligible when we observe that the dissipative flux  (\ref{se13}) is constant. Instead, for the second family of solutions the temperature (\ref{T2}) clearly exhibits the effects of transient phenomena (i.e. those  depending on $\tau$).

Finally we would like to conclude with a general comment: all the models exhibited above were  found with the sole purpose to illustrate the richness of solutions endowed with hyperbolical symmetry and sharing the general physical and geometrical properties (excluding the isometry group) characterizing the LTB space--times. It is now up to cosmologists and astrophysicists to decide if any of the above models (or any other HSLTB non described in this manuscript) could be of any use in the study of  specific scenarios, as for example cosmological models beyond the  standard FRW solution \cite{cm1, cm2}.

\begin{acknowledgments}
This work was partially supported by Ministerio de Ciencia, Innovacion y Universidades. Grant number:
PGC2018096038BI00, and Junta de Castilla y Leon. Grant number: SA096P20.
LH wishes to thank  Universitat de les  Illes Balears  and Departamento de F\'isica Fundamental at Universidad de Salamanca, for financial support and hospitality. ADP  acknowledges hospitality of the
Departament de F\'isica at the  Universitat de les  Illes Balears.

\end{acknowledgments}

\end{document}